\documentclass[aps,pre,reprint,superscriptaddress,longbibliography,showpacs,10pt]{revtex4-1}

\usepackage{color}
\usepackage{amsmath}
\usepackage{amssymb}
\usepackage{wrapfig}
\usepackage{braket}
\usepackage{amsthm}
\usepackage{natbib}
\usepackage{dsfont}
\usepackage{graphicx}
\usepackage[colorlinks = true, citecolor = blue, linkcolor = blue, urlcolor=blue]{hyperref}

\definecolor{green}{rgb}{0.0, 0.5, 0.0}

\definecolor{blue}{rgb}{0.0, 0.0, 0.5}

\begin{document}
\title{Metastability in an open quantum Ising model}

\author{Dominic C. Rose}
\affiliation{School of Physics and Astronomy, The University of Nottingham, Nottingham, NG7 2RD, United Kingdom}

\author{Katarzyna Macieszczak}
\affiliation{School of Physics and Astronomy, The University of Nottingham, Nottingham, NG7 2RD, United Kingdom}
\affiliation{School of Mathematical Sciences, University of Nottingham, Nottingham, NG7 2RD, United Kingdom}

\author{Igor Lesanovsky}
\affiliation{School of Physics and Astronomy, The University of Nottingham, Nottingham, NG7 2RD, United Kingdom}

\author{Juan P. Garrahan}
\affiliation{School of Physics and Astronomy, The University of Nottingham, Nottingham, NG7 2RD, United Kingdom}

\begin{abstract}
We apply a recently developed theory for metastability in open quantum systems to a one-dimensional dissipative quantum Ising model. Earlier results suggest this model features either a non-equilibrium phase transition or a smooth but sharp crossover, where the stationary state changes from paramagnetic to ferromagnetic, accompanied by strongly intermittent emission dynamics characteristic of first-order coexistence between dynamical phases. We show that for a range of parameters close to this transition/crossover point the dynamics of the finite system displays pronounced metastability, i.e., the system relaxes first to long-lived metastable states, before eventual relaxation to the true stationary state.  From the spectral properties of the quantum master operator we characterise the low-dimensional manifold of metastable states, which are shown to be probability mixtures of two, paramagnetic and ferromagnetic, metastable phases.  We also show that for long times the dynamics can be approximated by a classical stochastic dynamics between the metastable phases that is directly related to the intermittent dynamics observed in quantum trajectories and thus the dynamical phases.
\end{abstract}

\date{\today}

\maketitle

\section{Introduction}
Through the use of experimental platforms such as ultracold atomic gases, trapped ions and Rydberg atoms a wide range of non-equilibrium phenomena in open quantum many-body systems have now been realised \cite{Pritchard2010,Blatt2012,Britton2012,Dudin2012,Peyronel2012,Gunter2013}. This has been accompanied by a significant increase in the theoretical understanding of such systems; numerically, through the use of techniques adapted from atomic-optics and closed quantum systems \cite{Daley2014ManyBodyQJMC}, and analytically via field theoretical methods \cite{Sieberer2015KeldyshReview,Maghrebi2016}, among other approaches.

One question of interest addressed by theoretical studies of non-equilibrium open many-body quantum systems has been the understanding of the phase structure of the stationary state, given the possibility that these systems display phase transitions---singular changes in the stationary state in the limit of large system size---which may be distinct from those of equilibrium systems.  The aim of many of such studies has been to characterise the phase structure of non-equilibrium steady states via static order parameters, such as the magnetisation in spin systems, and spatial correlation functions more generally, see e.g.~\cite{{Diehl2013,Sieberer2015KeldyshReview,Maghrebi2016}}.  
However, even in the case where such transitions are not present, the dynamics of these systems can be very rich and cooperative.  One way to uncover this dynamical complexity via large-deviation methods (such as the so-called ``thermodynamics of trajectories'', see e.g.~\cite{{Juan2010ThermoTraj,Igor2013Characterization}}, or the related full-counting statistics, see e.g.~\cite{Esposito2009}), which allow to  define and characterise dynamical phases where order parameters are dynamic time-integrated observables, such as total count of emissions into the environment.

\begin{figure}[ht!]
\includegraphics[width=1\columnwidth]{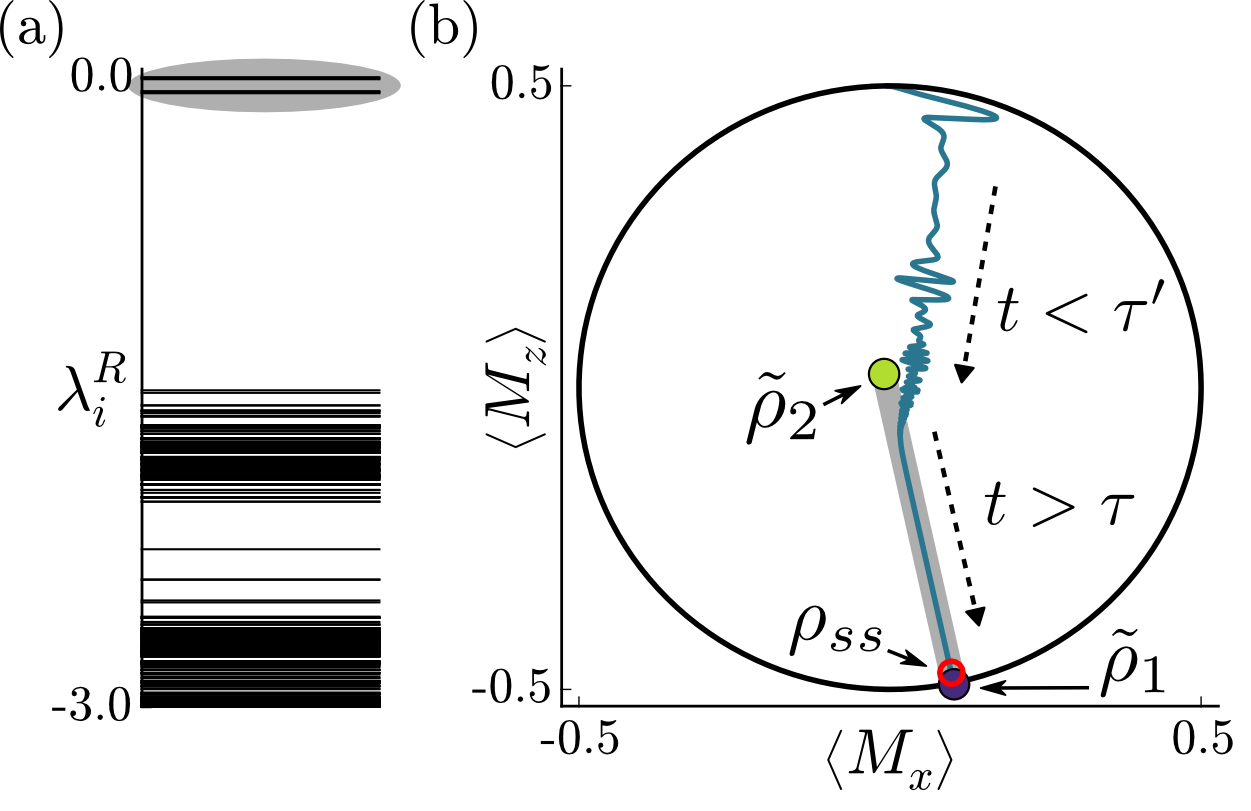}
\caption{\label{SpectrumToMM}(a) Real part of the spectrum of the master operator  of the open quantum Ising model defined in Sec.~\ref{OpenIsing}, for size $N=7$, with spin interaction $V/\kappa=250$ and transverse field $\Omega/\kappa=50$, where $\kappa$ is decay rate. Spectrum is plotted in units of $\kappa$. The two eigenvalues with largest real parts, ${\lambda}_{1,2}^{R}$, highlighted in grey, correspond to the only eigenmodes that contribute to the dynamics at long times. The large gap to the next eigenvalue, ${\lambda}_{3}^{R}$,  causes metastability in the dynamics. (b) The metastable manifold (MM) induced by the spectral structure of (a). System states are represented by the average total magnetisation per spin in the $x$ and $z$ directions. The MM (grey line) is bounded by two extreme metastable states $\tilde{\rho}_{1}$ (purple dot) and $\tilde{\rho}_{2}$ (green dot). The evolution from an initial state $\rho_\uparrow$ of all spins up (blue curve), undergoes a fast evolution for $t\ll\tau'=-1/{\lambda}_{3}^{R}$ onto the MM. After a long period of apparent stationarity ($\tau'\ll t\ll\tau=-1/{\lambda}_{2}^{R}$), the states on the MM evolve towards the true stationary state $\rho_{\rm ss}$ (red circle) for $t\gg\tau$.}
\end{figure}

In this paper we focus on a different aspect of open quantum many-body systems, that of {\em metastability} of the dynamics.  Metastability is a characteristic feature of the dynamics of slow relaxing systems, namely, the partial relaxation into long-lived states before eventual decay to the true stationary state.  This is a consequence of separation of timescales in the dynamics. In such systems the transient dynamics will appear non-ergodic, in the sense that different initial conditions will relax to different metastable states, before ergodicity is eventually restored. 
In classical systems, metastable behaviour typically occurs in two distinct situations:  one is in the dynamics near a (usually first-order) static phase transition \cite{Chaikin2000}, where metastable dynamics is intimately related to the existence of static phases; the second one is in glass forming systems \cite{Binder2011} where metastable behaviour is not obviously related to any static features (see e.g.~\cite{Biroli2013} for various viewpoints on this issue).  

Here we focus on the first instance above for the case of a many-body quantum system subject to dissipation where metastability can be traced back to distinct features of its stationary properties, 
a transverse field Ising model with on-site dissipative decay 
\cite{Lee2012CollectiveRydberg,Ates2012Ising,Weimer2015,Weimer2015b,Maghrebi2016,Overbeck2016,Matteo2016}.  We study the dynamics of this model by means of a recent theory for metastability in open quantum systems \cite{Kasia2016Meta}.  This new approach is a generalisation to open quantum systems,  evolving according to a Lindblad master equation \cite{Lindblad1976Equation,Breuer2002OpenQuant}, of the method of Refs.~\cite{{GaveauSchulman1987Meta,Gaveau1998,Bovier2002,Gaveau2006,Kurchan2009}} for classical Markovian systems.  This approach relies on the fact that metastable behaviour necessarily results from a separation of eigenmodes in the spectrum of the master operator generating the dynamics, thus allowing one to investigate both the structure of the metastable states and the long-time dynamics of the system using a reduced number of parameters \cite{Kasia2016Meta}. The metastable states form a metastable manifold (MM), a convex subset of the system states (see Fig.~\ref{SpectrumToMM}, details in Sec.~\ref{MetaTheory}) on which the long-time dynamics takes place \cite{Kasia2016Meta}.  In Ref.~\cite{Kasia2016Meta} this approach was applied to few-body models more related to problems in quantum information.  Here we demonstrate the effectiveness of this approach 
for a many-body system where metastability emerges cooperatively with increasing system size. The dissipative Ising model we consider is also of particular interest as it can be realised experimentally through the use of Rydberg atoms in optical lattices \cite{{Lee2012CollectiveRydberg,Ates2012Ising}}. 

The dissipative quantum Ising model we study here has the basic ingredients of a system where metastability is a dynamical consequence of static features.  In order to clarify what we mean, consider for example a classical ferromagnet at low temperatures in the thermodynamic limit~\cite{Chaikin2000}.  For such a system, if one starts close to the coexistence point (say at small positive magnetic field) but in a state belonging to the wrong phase (in this case with initial negative magnetisation) there will be an initial fast relaxation within this phase, before a much longer relaxation to the eventual equilibrium state within the stable phase (of positive magnetisation).  This occurs due to the existence of a large, but finite, free-energy barrier that needs to be crossed from the metastable phase to the stable phase.  Hysteretic behaviour is a manifestation of this phenomenon.  Furthermore, zero field  is the point of a first-order transition, and there is strict coexistence between the phases of positive and negative magnetisation, which means that each is stable at long times, giving rise to ergodicity breaking. 
It is important to note that such a transition only occurs in the thermodynamic limit.  For finite size, even at zero field, dynamics is metastable, with the metastable states being those of non-zero magnetisation, and the true stationary state that of paramagnetic equilibrium.  Dynamics in this case will also be intermittent displaying many ``magnetisation reversals'' \cite{Chaikin2000}.  In this sense, metastability which has its origin in distinct static ``phases'' does not require the presence of a strict phase transition, only that the ``phases'' are long lived and kinetically only weakly connected (in the example of the ferromagnet, separated by a free-energy barrier that is large compared to temperature).  In fact, usually the simplest way to anticipated such structure is from a mean-field approximation, even if there is no singular transition in the exact problem.

Previous work~\cite{Ates2012Ising} on the dissipative quantum Ising model suggested a bistable character to the stationary state of this system.  Within a mean-field approximation~\cite{Ates2012Ising} this model has a region of its parameter space in which it exhibits a strict first-order coexistence, terminating at a critical point, between ferromagnetic and paramagnetic steady states.  Related to this, dynamics of this model in one dimension for finite size was shown to display pronounced intermittency between periods of photon emission activity and inactivity in quantum jump trajectories~\cite{Ates2012Ising}.  While the mean-field approximation is not necessarily expected to be accurate even in the thermodynamic limit in one-dimension \cite{Weimer2015,Weimer2015b,Maghrebi2016}, it does provide an idea of the basic phase structure of the model.  As in the case discussed above of a classical ferromagnet, for metastability to be present all that is required is that states with distinct static character are approximate stationary states of the dynamics.  Using the methods of \cite{Kasia2016Meta} we show the relation between the ferromagnetic/paramagnetic states that mean-field suggest and the metastable states of the dynamics of the dissipative quantum Ising model at finite size.  We show that the short time dynamics relaxes to the metastable manifold in a way that is dependent on initial states (i.e., apparent but transient non-ergodicity).  We further show that the long-time evolution within the MM is that of an effective classical stochastic dynamics, even though the underlying many-body system features quantum effects.  This is in line with recent field-theoretic studies of open spin lattices, which show the long time properties of similar models are often classical \cite{Maghrebi2016,Sieberer2015KeldyshReview}.  

The paper is organized as follows. In Sec.~\ref{MetaTheory} we briefly review the spectral approach to metastability in open quantum systems, along with the effective long-time description, giving a particular focus to the case when the MM is one-dimensional as is the case for the dissipative quantum Ising model. 
Section~\ref{OpenIsing} is the central, providing an in-depth analysis of metastability in the model.  We begin in Subsec.~\ref{TDMF} by returning to the mean-field approach used in~\cite{Ates2012Ising}, but focusing on time-evolution instead of only steady state(s), and discussing the effect of neglecting fluctuations from spatial correlations.  We then apply the spectral approach of Ref.~\cite{Kasia2016Meta} to the exact model of $N=7$ spins, and find a parameter regime where the system is metastable (Subsec.~\ref{SpectAnalysis}). This is followed by characterisation of the MM in terms of the properties of metastable phases (Subsec.~\ref{eMSCharact}). In Subsec.~\ref{ExpandCorr} we show how metastability can be observed in the behaviour of expectation values and autocorrelations of system observables. Finally, we demonstrate how the metastability in the spin magnetisation and the effective dynamics within the MM manifests in individual realisations of the system evolution through intermittency in activity of quantum trajectories (Subsec.~\ref{Intermittence}).

\section{Metastable manifolds and effective evolution}\label{MetaTheory}

In this section we review the theory for metastability in open quantum systems with Markovian dynamics introduced recently in Ref.~\cite{Kasia2016Meta}.  Evolution of a system state described by the density matrix $\rho$ is generated by the Lindblad master operator
\begin{equation}\label{LindbladOp}
\mathcal{L}(\rho) = -i[H,\rho]+\sum_{j}\left[{J}_{j}\rho{J}_{j}^{\dagger}-\frac{1}{2}\{{J}_{j}^{\dagger}{J}_{j},\rho\}\right],
\end{equation}
i.e. $d\rho/dt=\mathcal{L}(\rho)$ \cite{Breuer2002OpenQuant}. Here $H$ is the Hamiltonian of the system and the jump operators ${J}_{j}$ mediate the system-bath interaction, providing coupling of the system to the surrounding environment. If these interactions lead to emissions of a quanta of energy, e.g. photons being emitted by atoms coupled to the vacuum electromagnetic field, the action of jump operators can be detected through continuous measurements, e.g. photon counting~\cite{Ates2012Ising}. 

Since this operator acts linearly on the density matrix, the evolution it generates can be understood in terms of its eigenvalues and eigenmatrices \cite{JordanBlock}. Let the eigenvalues of the master operator be  ${\lambda}_{k}={\lambda}_{k}^{R}+i\,{\lambda}_{k}^{I}$, with ${\lambda}_{k}^{R}$ and ${\lambda}_{k}^{I}$ representing the real and imaginary parts, respectively. These are ordered such that ${\lambda}_{k}^{R}\geq{\lambda}_{k+1}^{R}$. Note that the real parts are necessarily less than or equal to $0$,  with $0$ corresponding to steady states, due to the (completely) positive trace-preserving evolution generated by the master operator, Eq.~\eqref{LindbladOp}. Let the left and right eigenmatrices (note that $\mathcal{L}$ is not hermitian in general) associated to ${\lambda}_{k}$ be denoted ${L}_{k}$ and ${R}_{k}$ respectively. We choose normalization s.t. $\text{Tr}({L}_{k}{R}_{k'})={\delta}_{kk'}$. Generically, the steady state ${\rho}_{\rm ss}$ is unique (as is the case for the finite size dissipative Ising chain discussed in this work) and we further normalize it s.t. $\text{Tr}({\rho}_{\rm ss})=1$ and thus also $L_1=\mathds{1}$. Defining ${c}_{k}=\text{Tr}[\rho(0){L}_{k}]$, for an initial system state $\rho(0)$, the system state at time $t$ is given by
\begin{equation}\label{Expansion}
\rho(t)={e}^{t\mathcal{L}}[\rho(0)]={\rho}_{\rm ss}+\sum_{k}{c}_{k}{e}^{t{\lambda}_{k}}{R}_{k}.
\end{equation}

We are interested in metastable behaviour. Therefore we assume there is a large separation in the values of the real parts of the eigenvalues, ${\lambda}_{m}^{R}\gg{\lambda}_{m+1}^{R}$, cf.\ Fig.~1(a), and denote the corresponding time scales by ${\tau}=-1/{\lambda}_{m}^{R}$ and ${\tau}'=-1/{\lambda}_{m+1}^{R}$. If we consider times $t\gg\tau'$ then we may neglect terms beyond the $m$-th in the sum in Eq.~\eqref{Expansion}. For $t\ll\tau$, we also have ${e}^{t{\lambda}_{k}}\approx 1$ for $k\leq m$, and the reduced expansion defines metastable states of the system, see Fig.~\ref{SpectrumToMM}. Thus, the projection $\mathcal{P}$ onto the MM is given by $\mathcal{P}\rho=\text{Tr}(\rho){\rho}_{\rm ss}+\sum_{k=2}^{m}{c}_{k}{R}_{k}$. Furthermore, note that the MM is determined by a subset in ${\mathbb{R}}^{m-1}$ given by the coefficients $\{{c}_{k}\}_{k=1}^m$, whose values are bounded by the maximum and minimum eigenvalues of the relevant left eigenmatrices.  For times $t\gg\tau$, only first $m$ modes contribute to the system dynamics, cf.\ Eq.~\eqref{Expansion}. Therefore dynamics takes place essentially only inside the MM and is effectively generated by the projection of the master operator on the MM, $\mathcal{L}_{\text{eff}}=\mathcal{P}\mathcal{L}\mathcal{P}$.

\subsection{Metastable manifold}

The general structure for MMs is discussed in~\cite{Kasia2016Meta}.  We will see in Sec.~\ref{SpectAnalysis} that for the open quantum Ising model the MM is 1-dimensional, that is, $m=2$, which is the case we now restrict to. The basis for the MM is given by ${\rho}_{\rm ss}$ and ${R}_{2}$, but ${R}_{2}$ lacks interpretation as a state of the system since $\text{Tr}({R}_{2})=0$ due to trace-preserving dynamics. Instead, we can make use of the convexity of the MM (inherited from the convexity of the full space of density matrices) to consider metastable states as a linear combination of the states at either end of the MM, which we call {\em extreme metastable states} (eMSs)~\cite{Kasia2016Meta},
\begin{eqnarray}
\tilde{\rho}_{1}&=&{\rho}_{\rm ss}+{c}_{2}^{\rm max}{R}_{2},\label{eMS1}\\ \tilde{\rho}_{2}&=&{\rho}_{\rm ss}+{c}_{2}^{\rm min}{R}_{2},\label{eMS2}
\end{eqnarray}
where ${c}_{2}^{\rm max}$ and ${c}_{2}^{\rm min}$ are the maximum and minimum eigenvalues of the Hermitian ${L}_{2}$, respectively.

Writing $\mathcal{P}\rho={p}_{1}\tilde{\rho}_{1}+{p}_{2}\tilde{\rho}_{2}$ and using ${p}_{1}=1-{p}_{2}$ (from convexity of the MM), the coeffcients $p_{1,2}$ are given by the expectation value, $p_{1,2}=\mathrm{Tr}(\tilde{P}_{1,2}\rho)$, of the observables
\begin{eqnarray}
\tilde{P}_{1}&=&({L}_{2}-{c}_{2}^{\rm min}I)/{\Delta}{c}_{2},\label{P1}\\
\tilde{P}_{2}&=&(-{L}_{2}+{c}_{2}^{\rm max}I)/{\Delta}{c}_{2}, \label{P2}
\end{eqnarray}
where ${\Delta}{c}_{2} = {c}_{2}^{\rm max} - {c}_{2}^{\rm min}$. 

As $\tilde{\rho}_{1}$, $\tilde{\rho}_{2}$ are the eMSs, the coeffcients are necessarily positive, $p_{1,2}\geq 0$, leading to the interpretation of any state on the MM as a probabilistic mixture of the two eMSs. Thus, $\tilde{\rho}_{1}$ and $\tilde{\rho}_{2}$ constitute two metastable phases and one can understand the properties of all metastable states by characterising the properties of only these two, as done in Sec.~\ref{eMSCharact}. 
For a basis of any other two states on the MM, one of the coefficients would necessarily be negative for certain parts of the MM.

\subsection{Effective dynamics within metastable manifold}

The evolution for times $t\gg\tau$ effectively takes place on the metastable manifold towards the steady state~\cite{Kasia2016Meta}. Its generator $\mathcal{L}_{\text{eff}}=\mathcal{P}\mathcal{L}\mathcal{P}$, can be now expressed in the basis of the eMSs, cf.\ Eqs.~(\ref{eMS1}-\ref{P2}), 
\begin{equation}\label{EffectiveMasterOp}
\mathcal{L}_{\text{eff}}=\frac{{-\lambda}_{2}}{\Delta{c}_{2}}
\begin{pmatrix}
-{c}_{2}^{\rm max} & -{c}_{2}^{\rm min}\\
{c}_{2}^{\rm max} & {c}_{2}^{\rm min}
\end{pmatrix},
\end{equation}
yielding the dynamics of the probabilites ${p}_{1}$, ${p}_{2}$ as  $d\textbf{p}/dt={\mathcal{L}}_{\text{eff}}\,\textbf{p}$, where $\textbf{p}^{T}=({p}_{1},{p}_{2})$. Note that the entries in each column of $\mathcal{L}_{\text{eff}}$ add to $0$, which corresponds to probability conservation. Moreover, since ${\lambda}_{2}<0$ and ${c}^{\rm max}_{2}>0$, ${c}^{\rm min}_{2}\leq0$ [as $\text{Tr}({L}_{2}{\rho}_{\rm ss})=0$ by orthogonality of eigenmodes], the diagonal elements are negative and the off-diagonal positive, which ensures positivity of the dynamics. Thus, $\mathcal{L}_{\text{eff}}$ is a generator of classical stochastic dynamics between two metastable phases, $\tilde{\rho}_{1}$ and $\tilde{\rho}_{2}$. 

Note that the dynamics induced by $\mathcal{L}_{\text{eff}}$ between $\tilde{\rho}_{1}$ and $\tilde{\rho}_{2}$ satisfies the detailed balance, with the exit time from a metastable phase given by the inverse of the diagonal entry, e.g. $(-{c}_{2}^{\rm max}{\lambda}_{2})^{-1}=\tau\,({c}_{2}^{\rm max})^{-1}$ for $\tilde\rho_1$, and the stationary probability determined by the ratio of the off-diagonal terms, $-{c}_{2}^{\rm min}/{c}_{2}^{\rm max}$, independent from $\tau$.

\begin{figure}
\includegraphics[width=0.8\columnwidth]{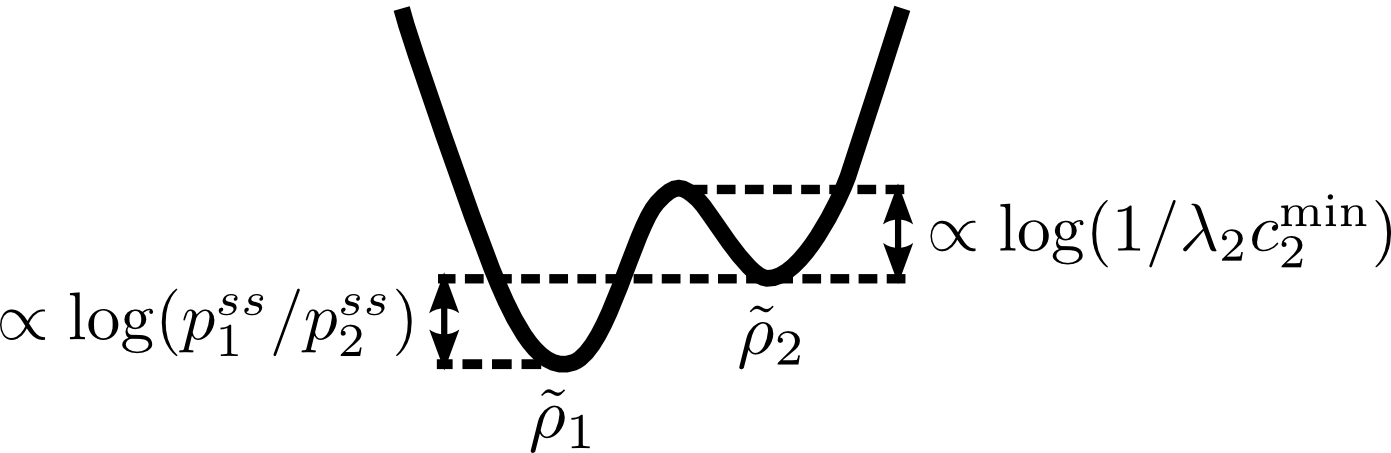}
\caption{
\label{wells} The Sketch for the two-basin analogy of metastability when the metastable manifold is one dimensional. The figure represents a one-dimensional Landau free-energy. The two eMSs are analogous to the two free-energy basins corresponding to two phases of the system, and their free-energy difference is analgous to the logarithm of their relative probabilities. In the thermodynamic case, the timescale for connecting the basins is the exponential of the free-energy barrier (divided by temperature -- the usual Arrhenius law for thermal escape). In this sense, since in our dynamical case the timescale is controlled by the gap, the log of the gap is analogous to the barrier. See main text for discussion. for connecting the basins, which in the classical case is determined by the free-energy barrier (typically as an exponential of the free-energy barrier over temperature), in our dynamical problem is determined by the gap in the master operator. See main text for discussion.
}
\end{figure}

This dynamics is analogous to the effective long-time dynamics in finite classical systems close to conditions where coexistence between two stable phases appears in the thermodynamic limit, such as for example a two-dimensional classical Ising model at low temperature in the ferromagnetic phase with small negative magnetic field favouring magnetisation down \cite{Chaikin2000}.  Intuitively, one can imagine such a situation in terms of motion in a Landau free-energy, a function of a collective variable such as magnetisation, which features two minima corresponding to the two competing phases, one of positive magnetisation and a favoured one of negative magnetisation, 
cf.\ Fig.~\ref{wells}.  The height of the barrier between the two wells is related to the surface tension of creating droplets of one phase in a background of the other.  In finite systems (or away from the coexistence point in the thermodynamic limit) this barrier can be overcome by thermal fluctuations. When temperature is low, such that thermal fluctuations are weak, an initial configuration of all spins down would initially relax via small scale fluctuations within the positive magnetisation phase (intuitively ``rolling down the hill'' towards its closest free-energy minimum) for $t\gg\tau'$.  At longer times, $t\sim\tau$ (dependent on the height of the barrier separating the wells, typically in an Arrhenius fashion for the case of thermal excitations), large enough domains of the spin-down phase are created allowing the system to overcome the free-energy barrier.  The subsequent switching dynamics between the two basins leads at long times to equilibration.  Note that due to detailed balance, the ratio of the occupation in the equilibrium state between the two wells depends on the difference of the free energies, not the barrier height which only sets the relaxation time.

\subsection{Observation of metastability}

Metastability of the system dynamics can be observed in the behaviour of expectation values and autocorrelations of system observables~\cite{Kasia2016Meta,Sciolla2015}. In Sec.~\ref{ExpandCorr} we focus on measuring a given spin magnetisation (or equivalently the probability distribution of the magnetisation). For long times $t\gg\tau'$ the exact dynamics should be captured by the effective long-time description, see Eq.~\eqref{EffectiveMasterOp}, and for the expectation value of an observable $O$ we arrive at
\begin{equation}\label{ApproxExp}
\left\langle{O}\right\rangle(t)\approx\textbf{o}^{T}\,\textbf{p}(t),
\end{equation}
where ${\textbf{o}}_{i}=\text{Tr}(O\tilde{\rho}_{i})$. Observation of the metastability in the expectation value requires preparation of an initial system state different from the steady state. Nevertheless, for the system in the steady state,  metastability can be observed in the time-autocorrelation of a system observable. When time between measurements is greater than $\tau'$, the state conditioned on the initial measurement relaxes into the MM and $\left\langle{O}(t){O}(0)\right\rangle=\text{Tr}(\mathcal{O}{e}^{t\mathcal{L}}\mathcal{O}{\rho}_{\rm ss})\approx\text{Tr}(\mathcal{O}{e}^{t\mathcal{L}_{\text{eff}}}\mathcal{P}\mathcal{O}{\rho}_{\rm ss})$, where $\mathcal{O}$ is the superoperator representing measurement of the observable $O$. As $\mathcal{P}{\rho}_{\rm ss}={\rho}_{\rm ss}$, we have 
\begin{equation}\label{ApproxCor}
\left\langle{O}(t){O}(0)\right\rangle\approx\textbf{o}^{T}\,{e}^{t\mathcal{L}_{\text{eff}}}\,{\textbf{O}}\,\textbf{p}_{\rm ss},
\end{equation}
where ${\textbf{O}}_{ij}={\left(\mathcal{P}O\mathcal{P}\right)}_{ij}=\textrm{Tr}[\tilde{P}_{i}O({\tilde{\rho}}_{j})]$  and $\textbf{p}_{\rm ss}^T=(-{c}_{2}^{\rm min}/{\Delta}{c}_{2},{c}_{2}^{\rm max}/{\Delta}{c}_{2})$ is the stationary probability of $\rho_{\rm ss}$ between the metastable phases.

\begin{figure*}
\includegraphics[width=1\linewidth]{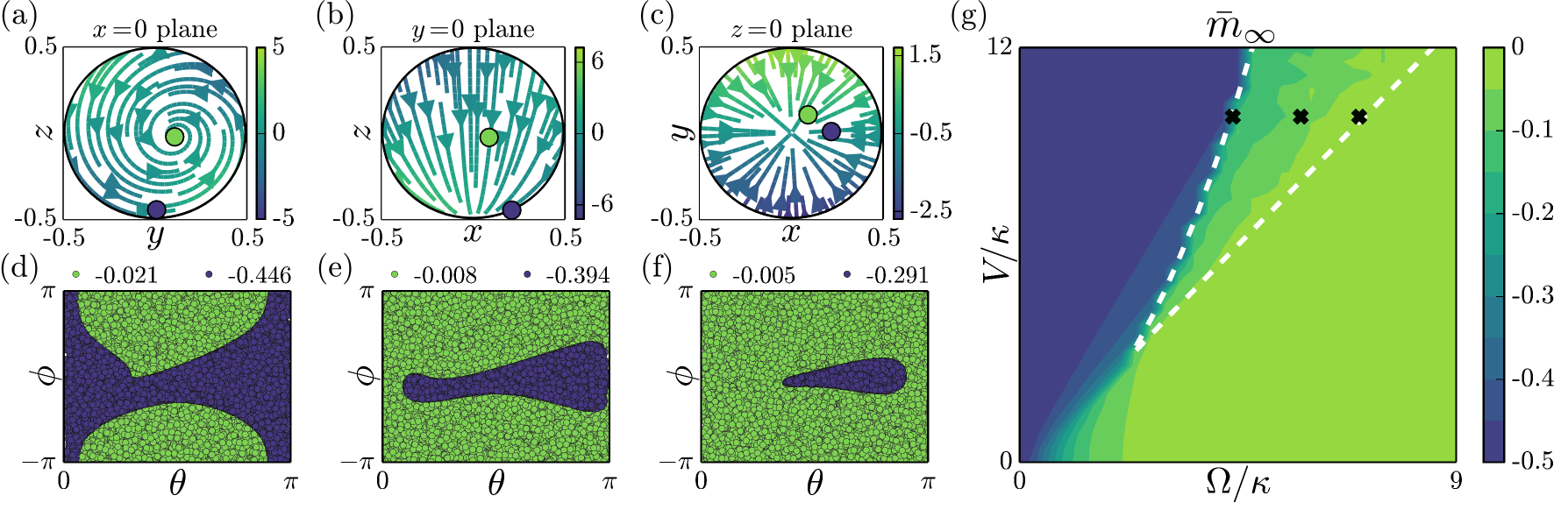}
\caption{\label{MeanField}(Color online) (a-c) Flow induced on cross sections of the Bloch ball by the dynamics in Eq.~(\ref{MeanFieldEqns}), for parameters $V/\kappa=10$, $\Omega/\kappa=4.4$ where the system is bistable. The colour of the line indicates the magnitude of the component perpendicular to the plane. The green and purple dots correspond to the two stationary solutions of the mean-field approximation projected onto the planes through the Bloch sphere. (d-f) Basins of attraction of the two mean-field stationary solutions, with the indicated values of $\left\langle{M}_z\right\rangle$, at three points in the bistable regime with parameters $V/\kappa=10$ and $\Omega/\kappa=4.4, 5.8, 7.0$, respectively, marked by black crosses in panel (g). (g) Mean-field phase diagram: $\overline{m}_{\infty}$, as a function of model parameters.  Spinodal lines separating one-solution from two-solution regions are 
indicated by white dashed lines.}
\end{figure*}

\section{Open quantum Ising model}\label{OpenIsing}

We now apply the theory~\cite{Kasia2016Meta} reviewed in the previous section to a dissipative quantum many-body system, an open version of the 1D transverse field Ising model, consisting of a chain of $N$ spin-1/2 particles with periodic boundary conditions coupled to a external bath. The Hamiltonian is given by
\begin{equation}\label{Hamiltonian}
H = \Omega\sum_{j=1}^{N}{S}^{(j)}_{x}+V\sum_{i=1}^{N}{S}^{(j)}_{z}{S}^{(j+1)}_{z},
\end{equation}
where $\Omega$ describes the transverse magnetic field and $V$ the interaction between neighbouring spins. The spin operators ${S}^{(j)}_{\alpha}=\frac{1}{2}{\sigma}^{(j)}_{\alpha}$, where $\alpha=\{x,y,z\}$. The jump operators mediating the  interaction with the bath couple to individual spins,
\begin{equation}\label{Jumps}
{J}_{j}=\sqrt{\kappa} \, {S}^{(j)}_{-}=\sqrt{\kappa} \, ({S}^{(j)}_{x}-i{S}^{(j)}_{y}),
\end{equation}
where the decay rate $\kappa$ is determined by strength of system-bath coupling, and $j=1,...,N$. This system can be realized using Rydberg atoms \cite{{Lee2012CollectiveRydberg,Muller2008TrappedRydberg}}, in which case the jumps can be seen to be photon emissions from individual spins in to the environment. The properties of these jump operators imply that the system dynamics are irreducible and the stationary state unique for all finite $N$ \cite{SchirmerWang2010}.

The order parameters we will consider are the magnetisations along each axis, 
\begin{equation}
{M}_{\alpha}=\frac{1}{N}\sum_{j=1}^{N}{S}_{\alpha}^{(j)}{.} \label{eq:M}
\end{equation}
We will find in Sec.~\ref{ExpandCorr} and \ref{Intermittence} the the $z$ component of the magnetisation in particular is sufficient to highlight metastable dynamics in this system. 

\subsection{Time dependent mean-field approximation}\label{TDMF}
We start our analysis by returning to the mean-field approximation studied previously in~\cite{Ates2012Ising}, this time however focusing on the dynamics rather than just the steady state. From the study of classical systems it is common for such an approach to give qualitatively the short time dynamics leading to the MM, since often long-time relaxation is non-perturbative and triggered by spatial fluctuations that are dismissed at the mean-field level, see e.g.~\cite{Binder2011}. 
We therefore expect to uncover within this approximation properties of the dynamics of the exact problem in its approach to the metastable regime ($t\ll\tau$).  In the next subsections we consider the whole dynamical regime by considering the full spectral properties of the exact problem. 

In this subsection we proceed as follows: we consider an approximation of the density matrix in the form of an arbitrary product state, i.e $\rho_N={\rho_1}^{\otimes N}$, where ${\rho}_{1}=\frac{1}{2}\mathds{1}+x{S}_{x}+y{S}_{y}+z{S}_{z}$. Note that here $x,y,z=\left\langle{M}_{x}\right\rangle,\left\langle{M}_{y}\right\rangle,\left\langle{M}_{z}\right\rangle$. These coefficients are time dependent, and we substitute this into the master operator, proceeding to trace out all but one spin, i.e. $d{\rho_1}/dt=\text{Tr}_{2,...,N}[\mathcal{L}(\rho)]$. This process gives a set of three non-linear differential equations on the coefficients given by
\begin{eqnarray}
\dot{x}&=&-\frac{\kappa}{2}x-2Vyz,\nonumber\\
\dot{y}&=&-\frac{\kappa}{2}y+2Vxz-\Omega{z},\label{MeanFieldEqns}\\
\dot{z}&=&-\frac{\kappa}{2}(1+2z)+\Omega{y}.\nonumber
\end{eqnarray}
The dynamics induced by these equations is highly oscillatory, with rotations about the $x$-axis induced by the transverse field $\Omega$, see Fig.~\ref{MeanField}\textcolor{blue}{(a)}, and rotations about the $z$-axis---weighted by the magnetisation $\left\langle{M}_{z}\right\rangle$, thus no rotation in the $z=0$ plane, Fig.~\ref{MeanField}\textcolor{blue}{(c)}---due to the spin-spin interaction with potential $V$. These oscillations are eventually damped by the decaying terms proportional to $\kappa$.\\

\emph{Stationary solutions}. The steady states of this system of equations are found from the solution of $\dot{x}=0$, $\dot{y}=0$ and $\dot{z}=0$, which gives rise to a 3rd order polynomial in $z$, see \citep{Ates2012Ising}. There is either one or two stable real solutions depending on the parameters chosen, with the set of parameters giving bistable dynamics lying between two ``spinodal'' lines, see Fig.~\ref{MeanField}\textcolor{blue}{(g)}. Outside this region the unique stationary state is ferromagnetic ($z\approx-0.5$) for small $\Omega$ and close to paramagnetic ($x,y,z$ small) for large $\Omega$. Inside the bistable region one of the two stationary states corresponds to a paramagnetic state while the other transitions smoothly from ferromagnetic to paramagnetic with increasing $\Omega$. The two steady states are shown on the flow diagrams in Fig.~\ref{MeanField}\textcolor{blue}{(a-c)}.

Since the mean-field approach is expected to capture the short time evolution we anticipate the two steady states to be qualitatively similar in properties to the eMS of the exact model. In order to compare the mean-field approximation with the exact dynamics during the metastable regime, we investigate the average asymptotic magnetisation, $\overline{m}_{\infty}$, reached from pure initial states $\rho_1$ sampled uniformly from the Bloch sphere, Fig.~\ref{MeanField}\textcolor{blue}{(g)}, i.e., the magnetisation of the steady state reached according to Eq.~(\ref{MeanFieldEqns}) (see also~\cite{Marcuzzi2014}).  Outside of the bistable regime this is simply the magnetisation of the unique mean-field steady state, $m_{\rm ss}$.  Within the bistable regime, $\overline{m}_{\infty}$ is the linear combination obtained from adding up the area of the basins of attraction for the two mean-field steady states, Figs.~\ref{MeanField}(d-f), multiplied by their respective magnetisations. 

For low $\Omega$ the basins are fairly even in size, giving a sharp transition at the left spinoidal line from $m_{\rm ss}\approx-0.5$ of the unique ferromagnetic steady state to $\overline{m}_{\infty}\approx-0.25$, cf.\ Fig.~\ref{MeanField}\textcolor{blue}{(d-f)}. As $\Omega$ increases the area of the basin for the steady state with a lower magnetisation shrinks continuously, along with the magnetisation of this state increasing, with the area of the basin vanishing at the right spinoidal line, leading to a smooth increase of $\overline{m}_{\infty}$ to $m_{\rm ss}\approx0$.  Below we consider the exact dynamics of a finite chain.  We will see that the  magnetisation, averaged over a uniform distribution of initial states, behaves during the metastable regime in a similar way to the mean-field result above.

\subsection{Spectral analysis of the master operator}\label{SpectAnalysis}

\begin{figure}
\includegraphics[width=1\linewidth]{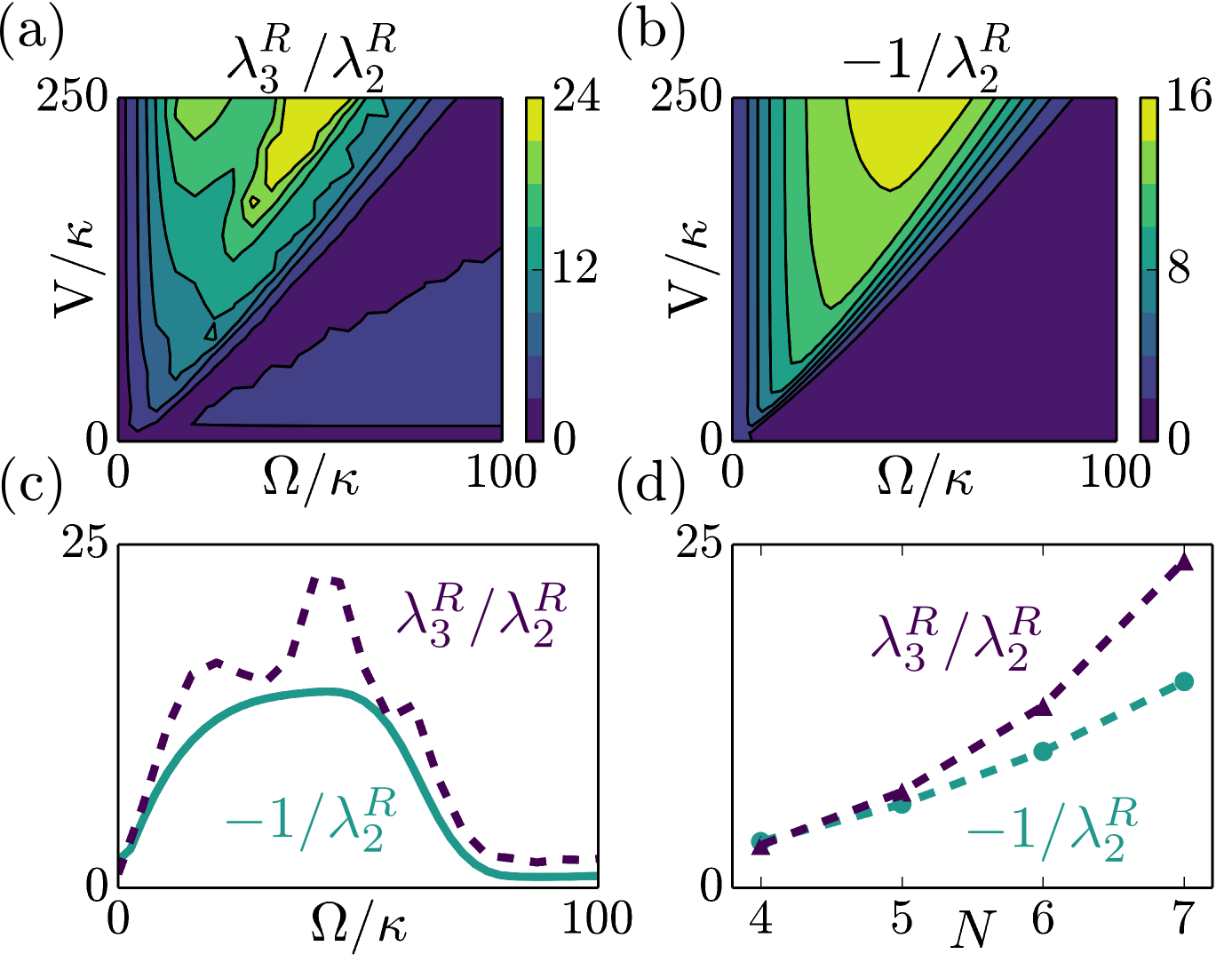}
\caption{\label{SpectrumData}(Color online) (a) The ratio of the real parts of the second and third eigenvalues ${\lambda}_{3}^{R}/{\lambda}_{2}^{R}$, plotted over  a region of the parameter space $0\leq{V}\leq250$ and $0\leq\Omega\leq100$. (b) The inverse gap $-1/{\lambda}_{2}^{R}$ plotted over the same region. (c) A cross section of plots (a) and (b) for $V=200$. The dashed (purple) line is the ratio while the solid (blue) line is the inverse gap. (d) The scaling of the maximum value of the ratio (purple triangles) and inverse gap (blue circles) in this range of parameters with system size.}
\end{figure}

Next we explore how the features of the mean-field treatment manifest in the exact master equation dynamics.  To this consider numerically a system with $N=7$ spins. We make use of the system translation invariance to first block diagonalize the master operator \cite{{BucaProsen2012LindSym,AlbertJiang2014LindSym}}.

For metastability we require separation in the real part of the spectrum of the Lindbladian, and we find that for this system this occurs  between the second and third eigenvalues, so that $m=2$  and the metastable manifold is 1D, see Figs.~\ref{SpectrumToMM}\textcolor{blue}{(a)} and \ref{SpectrumData}\textcolor{blue}{(b)}. The parameter region of metastable dynamics---light yellow wedge in Fig.~\ref{SpectrumData}\textcolor{blue}{(a)}---is suggestive of identification with the bistable region of the mean-field approximation, cf.\ Fig.~\ref{MeanField}\textcolor{blue}{(g)}, with a stretching to lower values of the transverse field. The separation in the spectrum is mainly due to the the spectral gap, $-{\lambda}_{2}^{R}$, getting small, Fig.~\ref{SpectrumData}\textcolor{blue}{(b,c)}. 
Metastability in the dissipative Ising model is a collective phenomenon, since the ratio ${\lambda}_{3}^{R}/{\lambda}_{2}^{R}$ increases with size, as does the inverse of the gap $-1/{\lambda}_{2}^{R}$, see Fig.~\ref{SpectrumData}\textcolor{blue}{(d)}. 

It remains an open question whether the gap strictly closes for some combination of parameters in the thermodynamic limit, as in the mean-field approximation.  This would correspond to a phase transition in the stationary state of the system.  What the mean-field approximation predicts is that this would be a transition between phases of different magnetisation, cf.\ the states at either side of the bistable region, Fig.~\ref{MeanField}\textcolor{blue}{(g)}.  Irrespective of whether there is a true transition in the thermodynamic limit, or a smooth but sharp crossover, the decrease of the gap with increasing size, Fig.~\ref{SpectrumData}\textcolor{blue}{(d)}, indicates that metastability is a consequence of the proximity to this transition/crossover point.  The system sizes accessible to numerics are far from large, therefore precluding a singular transition.  Dependence on size, however, indicates that such a change (whether a transition or sharp crossover) in the steady state magnetisation, ${m}_{\rm ss}=\text{Tr}({M}_{z}{\rho}_{\rm ss})$, from the ferromagnetic state at small~$\Omega$, $m_{\rm ss}\approx-1/2$, to the paramagnetic one at large~$\Omega$, $m_{\rm ss}\approx\text{Tr}({M}_{x}{\rho}_{\rm ss})\approx\text{Tr}({M}_{y}{\rho}_{\rm ss})\approx0$, should become very pronounced at large sizes. See Fig.~\ref{EMSChar}\textcolor{blue}{(a)}.

\subsection{Characterization of the metastable manifold}\label{eMSCharact}

\begin{figure}
\includegraphics[width=1\linewidth]{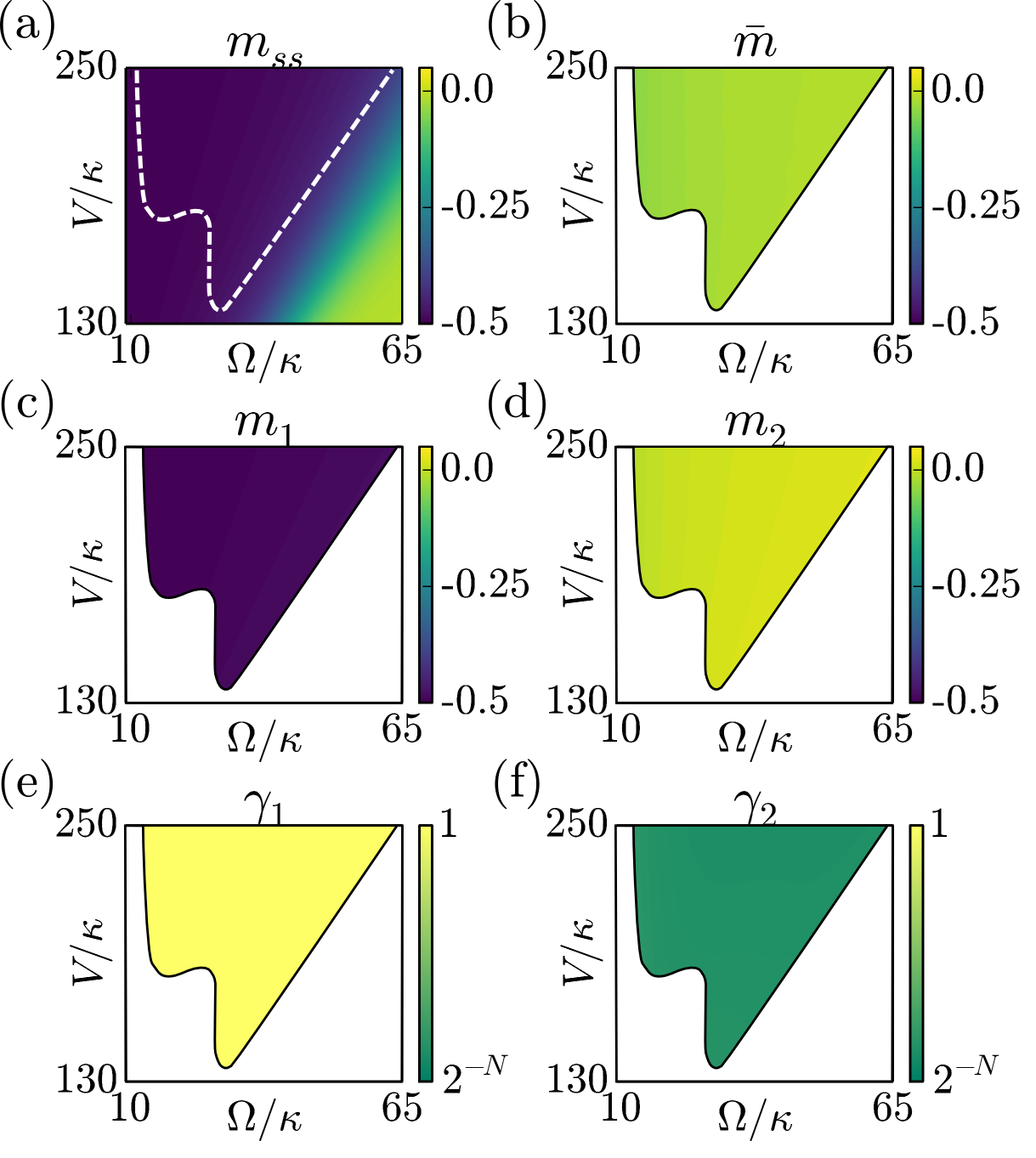}
\caption{\label{EMSChar}(Color online) (a) Magnetisation of the steady state ${m}_{\rm ss}$. The approximate boundary of the metastable region is indicated by the dashed white line. (b) The average $z$ magnetisation $\overline{m}$ in the metastable regime.  In this and the following panels we only show the area where metastability holds. (c-d) Expected $z$ magnetisations ${m}_{i}=\text{Tr}({M}_{z}\tilde{\rho}_{i})$ and (e-f) purities ${\gamma}_{i}=\text{Tr}(\tilde{\rho}_{i}^{2})$ of the two metastable phases.}
\end{figure}

The metastable manifold can be characterised by the properties of two eMSs defined by Eqs.~(\ref{eMS1},\ref{eMS2}), which we investigate below.  These two eMSs correspond to ferromagnetic (${m}_{1}\approx-0.5$) and paramagnetic (${m}_{2}\approx0$) states, see Fig.~\ref{EMSChar}\textcolor{blue}{(c-d)}. Note the similarity between 
the values of ${m}_{1}$ and ${m}_{2}$ resemble the magnetisations of two stationary states in the mean-field approximation, cf.\ Fig.~\ref{MeanField}\textcolor{blue}{(d-f)}.  

We consider now the behaviour of the magnetisation in $z$-direction, $\langle M_z\rangle$, during the metastable regime taken on average by pure initial states $|\psi\rangle$, see Fig.~\ref{EMSChar}\textcolor{blue}{(b)}, i.e. $\overline{m}=\overline{\mathrm{Tr}({M}_{z}\mathcal{P}|\psi\rangle\langle\psi|)}$, where the bar denotes average w.r.t.\ uniform sampling of $|\psi\rangle$. For most of the parameter regime where metastability is present we have $\overline{m}\approx0$, with a transition at small $\Omega$. This is analogous to $\overline{m}_{\infty}$ in the mean-field case, cf.\ Fig.~\ref{MeanField}\textcolor{blue}{(g)}. Here the size of the basins of attraction (given by $\frac{1}{N}\mathrm{Tr}\tilde{P}_1$ and $\frac{1}{N}\mathrm{Tr}\tilde{P}_2$, respectively) are imbalanced towards the paramagnetic phase for almost all of the metastable region, with a slight increase in area of the ferromagnetic phase at smaller fields. This suggests that the mean-field approximation captures the phenomenology of both the metastable phases and their basins of attraction. 

Note the contrast between the transition in $\overline{m}$ to the behaviour of the magnetisation $m_{\rm ss}$ of the steady state, cf.\ Fig.~\ref{EMSChar}\textcolor{blue}{(a)}, which takes place for large $\Omega$, outside the metastable parameter region. For the parameters where metastability is present, $\rho_{\rm ss}$  consists mostly of the ferromagnetic phase, $p_1^{\rm ss}\approx1$. This shows that the short time evolution leads on average to the vast majority of initial states evolving to the paramagnetic metastable phase on short time scales, while at longer times they decaying into the mostly ferromagnetic stationary state.

The two metastable phases differ also in purity $\gamma=\text{Tr}({\rho}^{2})$, see Fig.~\ref{EMSChar}\textcolor{blue}{(e-f)}, with the ferromagnetic phase being almost pure (${\gamma}_{1}\approx1$), in contrast the paramagnetic one almost completely mixed (${\gamma}_{2}\approx1/2^N$). This is a consequence of the two phases corresponding respectively to inactive and active periods in the photon emission records \citep{Ates2012Ising}. The fluctuations in photon emissions per unit time for the system in the paramagnetic phase cause it to be a mixture of many different pure states, i.e. the large number of jumps causes these periods in the trajectories to consist of many different states, while the relative inactivity of the ferromagnetic phases leaves the state approximately pure. The relation of the metastable phases to activity in quantum trajectories is further studied in Sec.~\ref{Intermittence}.

\subsection{Expectations, correlations and effective low-dimensional dynamics} \label{ExpandCorr}

\begin{figure}
\includegraphics[width=1\columnwidth]{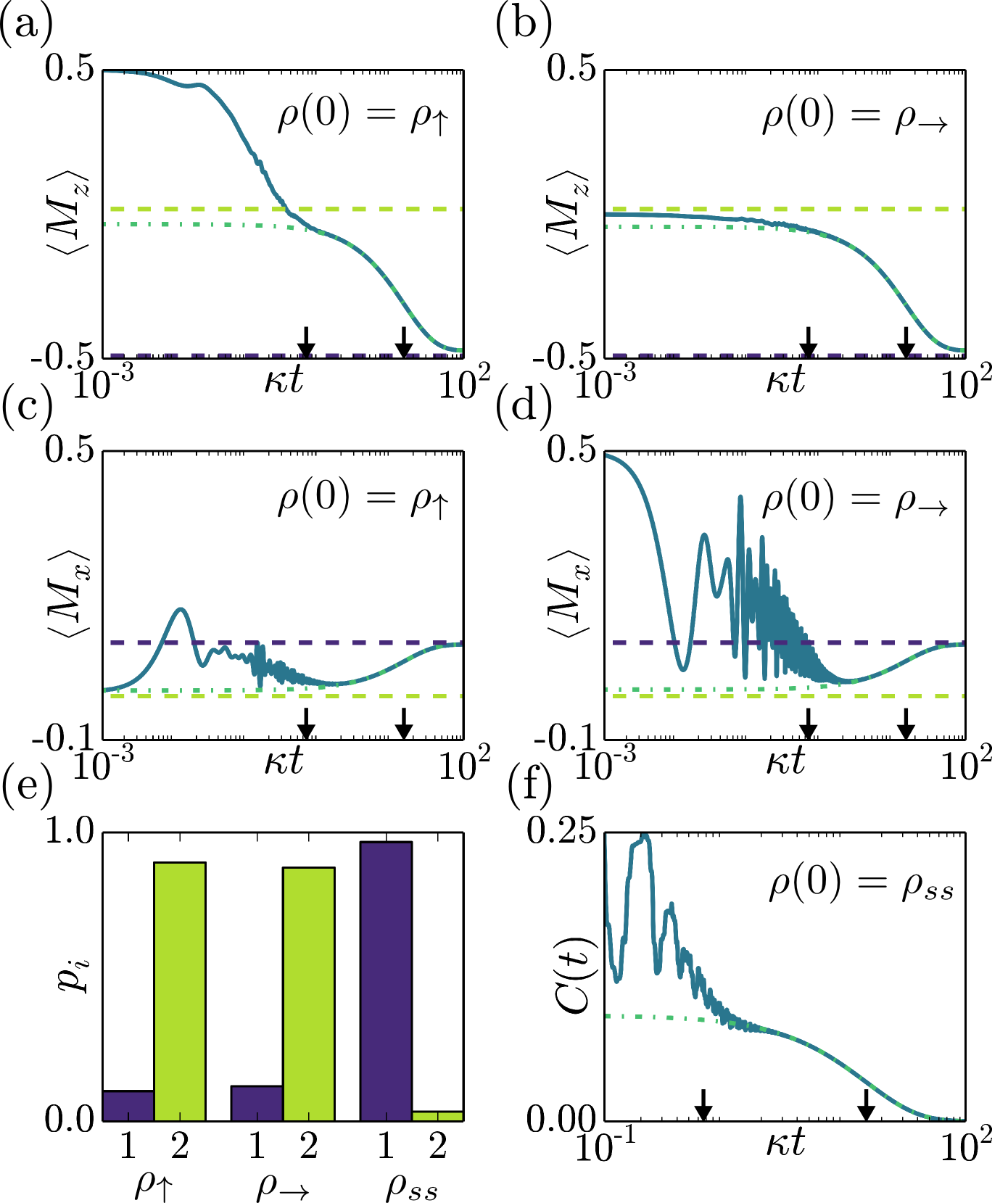}
\caption{\label{ObsMeta}(Color online) (a) Time evolution of the average $z$-magnetisation, $\left\langle{M}_{z}\right\rangle$ for the initial state $\rho_\uparrow$. The $z$-magnetisations of the metastable phases $\tilde{\rho}_{1,2}$ are represented by the dashed lines. (b) Same but for the initial state $\rho_\rightarrow$. 
(c,d) Time evolution of the average $x$ magnetisation, $\left\langle{M}_{x}\right\rangle$, for $\rho_\uparrow$ and $\rho_\rightarrow$. The average metastable values are given by dashed lines. (e) Histogram of the probability of finding $\rho_\uparrow$,  $\rho_\rightarrow$ and $\rho_{\rm ss}$ in each eMS during the metastable regime, see Eq. \eqref{P1}, \eqref{P2}. (f) Autocorrelation function of the projective measurement related to $M_z$ in $\rho_{\rm ss}$ (see the main text).
In panels (a-d,f) the time is in log scale, with the two arrows indicating the time scales $\tau'$ and $\tau$. The exact dynamics is plotted as solid curves, and the effective dynamics as dash-dotted lines. 
}
\end{figure}

After the initial relaxation the system state is approximately stationary during the metastable regime before eventual relaxation into the steady state. 
Metastability can be observed in the two-step decay of the expectation value $\left\langle{O}(t)\right\rangle$ and two-time correlation function $\left\langle{O}(t){O}(0)\right\rangle$ of an observable $O$ measured on the system. Here, we demonstrate that this is indeed the case when the observable is the magnetisation ${M}_{z}$, Eq.~\eqref{eq:M}.

We consider dynamics of two initial states: the ``all up" state ${\rho}_{\uparrow}={(\ket{\uparrow}\bra{\uparrow})}^{\otimes{N}}$, where ${\sigma}_{z}\ket{\uparrow}=\ket{\uparrow}$, and the ``all right" state ${\rho}_{\rightarrow}={(\ket{\rightarrow}\bra{\rightarrow})}^{\otimes{N}}$, where ${\sigma}_{x}\ket{\rightarrow}=\ket{\rightarrow}$. The expected values of the $z$-magnetisation, $\left\langle{M}_{z}(t)\right\rangle$, show a plateau in the metastable regime ($\tau'<t<\tau$), see Fig.~\ref{ObsMeta}\textcolor{blue}{(a-b)}, when the initial states evolve into mixtures of metastable phases, Fig.~\ref{ObsMeta}\textcolor{blue}{(e)}. Subsequent evolution takes the magnetisation to that of the mostly ferromagnetic stationary state at later times ($t>\tau$). Meanwhile, the expected value of the $x$-magnetisation, $\left\langle{M}_{x}\right\rangle$, Fig.~\ref{ObsMeta}\textcolor{blue}{(c-d)}, displays oscillations for times $t<\tau'$ suggestive of those present in mean-field model, cf.\ Fig.~\ref{MeanField}\textcolor{blue}{(a-c)}, which is expected to qualitatively represent the short time dynamics. The long-time behaviour is captured by the effective dynamics $\mathcal{L}_\text{eff}$, cf.\ Eq.~\eqref{ApproxExp}.  Both states evolve mostly into a paramagnetic metastable phase $\langle M \rangle\approx0$, despite being very different initially, $\mathrm{Tr}({\rho}_{\uparrow}{\rho}_{\rightarrow})=1/2^N$. This demonstrates the dominance of this metastable phase in the short time evolution.

In Fig.~\ref{ObsMeta}\textcolor{blue}{(f)} we show how metastability is also observed in the (time averaged) autocorrelation of a measurement conducted on the steady state. The correlations in the difference of probabilities of measuring magnetisations corresponding to the paramagnetic metastable phase, $P({M}_{z}=-N^{-1})+P({M}_{z}=N^{-1})$ as $N$ is odd, and the ferromagnetic one, $P(M=-1/2)$, feature a two-step decay, with large oscillations damped after short times and later dynamics simply given by Eq.~\eqref{ApproxCor}.

\subsection{Intermittent dynamics of quantum trajectories}\label{Intermittence}

\begin{figure*}
\includegraphics[width=1\linewidth]{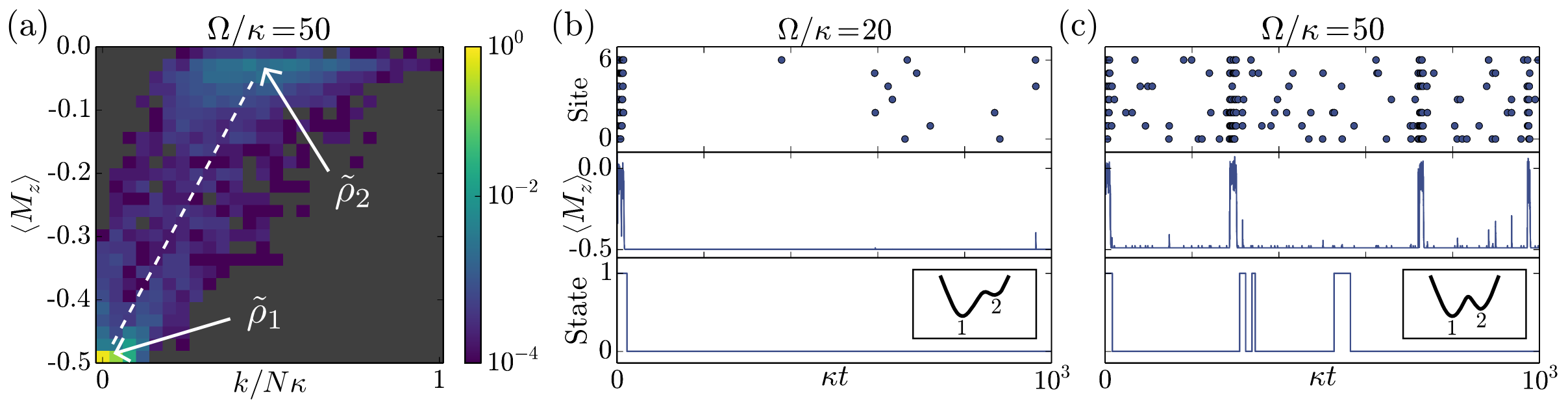}
\caption{\label{IntTraj} (Color online) (a) A 2d histogram of the average magnetisation $m$ and activity $k$ of time bins of length ${t}_{\text{bin}}=6\tau'$, for parameters $V/\kappa=250$, $\Omega/\kappa=50$ and $N=7$. The arrows indicate the expectation values for $\tilde{\rho}_{1}$ and $\tilde{\rho}_{2}$ respectively, with the decay $e^{{t}_{\text{bin}}\lambda_2}$ of $R_2$ taken into account. The white dashed line indicates the relation $\langle k\rangle /N \kappa = \langle M_z\rangle+\frac{1}{2}$. (b-c) Sample trajectories for two different sets of parameters in the metastable region, $V/\kappa=250$ with $\Omega/\kappa=20$, 50, for N=7 and time in linear scale. The first and seconds rows show the space and time resolved photon emissions and the corresponding magnetisation of the state over time respectively, for a sample QJMC trajectory. The third row shows a sample classical trajectory, jumping between the two metastable phases. The insets on the third row show sketches of the double well potentials of free energy leading to analogous classical evolution between the metastable states of two wells.  }
\end{figure*}

So far we have discussed the dynamics of the system described by the density matrix $\rho(t)$, which is governed by the master operator~\eqref{LindbladOp}, and the related dynamics of observables on the system, such as the spin magnetisation~\eqref{eq:M}. Instead, one can consider an unravelling of a master equation into quantum trajectories~\cite{Breuer2002OpenQuant}, which corresponds to a particular experimental realisation of system dynamics and a continuous measurement.  In this case, a quantum jump occurs when, say, the system jump operator $J_k$ flips the $k$-th spin from up-to down, cf.\ Eq.~\eqref{Jumps}, which is accompanied by the emission of a photon into the environment, which can be be detected. In this case, a measurement record up to time $t$ consists of information about which spins emitted photons and the emission times, see Fig.~\ref{IntTraj}\textcolor{blue}{(b-c)}.

Based on this record, a quantum trajectory---the conditional pure system state at any time $t'\leq t$---can be reconstructed~\cite{Breuer2002OpenQuant}. The density matrix $\rho(t)$ considered in earlier sections is simply an average over the ensemble of such pure conditional states at time $t$, i.e.\ over the system state at time $t$ in all realisations of the experiment. We now discuss how metastability is present not only in the master equation dynamics, but also in individual quantum trajectories. 

Photon emission records and quantum trajectories can be generated by the quantum jump Monte Carlo method (see e.g.~\cite{Daley2014ManyBodyQJMC}). The probability of observing no photon emission for time $t'$ for the system initially in $\ket{\psi}$  is given by $P_0(t')=\langle \psi(t')|\psi(t')\rangle$, where $\ket{\psi(t')}=e^{it'{H}_{\text{eff}}}\ket{\psi}$ with the effective Hamiltonian ${H}_{\text{eff}}=H-\frac{i}{2}\sum_{j=1}^N{J}_{j}^{\dagger}{J}_{j}$, cf.\ (\ref{Hamiltonian},\ref{Jumps}). The probability of observing a first photon emission from a $j$-th spin at time $t'$ is further given by ${P}_{j}(t')=\bra{\psi(t')}{J}_{j}^{\dagger}{J}_{j}\ket{\psi(t')}$, and the conditional pure system state is $\ket{{\psi}_{j}(t')}={J}_{j}\ket{\psi(t')}/\sqrt{{P}_{j}(t')}$. Using these definitions one can simulate the system evolution in terms of stochastic quantum trajectories \cite{Daley2014ManyBodyQJMC}.

The total number  of a photons emitted up to time $t$ for dynamics with a unique steady state obeys the Central Limit theorem~\cite{Guta2011,Catana2015}. Moreover, both the mean and the variance of the total number of emissions asymptotically ($t\gg\tau$) scale linearly in with time $t$, (for the case of the dissipative Ising model see~\cite{Ates2012Ising}). We argue that this also holds in the metastable regime for each of the metastable phases. As the total emission number $\Lambda(t)$ is an observable integrated over the observation time, for time $t$ well inside the metastability regime, the leading terms in its mean and variance are given by the contribution when the system state (averaged over trajectories) is metastable. Since the instantaneous rate $\mu$ of photon emissions from the system in a state $\rho$ is given by its magnetisation, $\mu=\sum_{j=1}^N \mathrm{Tr} ({J}_{j}^{\dagger}{J}_{j}\rho)=N\kappa\left(\mathrm{Tr} ({M}_{z}\rho)+\frac{1}{2}\right)$, the mean number of photons emitted in time $\tau'\ll t\ll \tau$ can be approximated by $\langle\Lambda(t)\rangle\approx t\,N\,\kappa \,(p_1 m_1+p_2 m_2 +\frac{1}{2})$. 

Consider the case of an initial state evolving into just one of the metastable phases.  In that case the mean activity, $k(t)=\Lambda(t)/t$, of trajectories is given by the magnetisation in the corresponding eMS, $\langle k(t)\rangle\approx N\kappa (m_1+\frac{1}{2})$ or $N\kappa(m_2+\frac{1}{2})$. As the correlations of dynamics inside a metastable phase can decay at most at times of order $\tau'$, in this case also the variance, $\Delta^2\Lambda(t)$, scales linearly in time for $t$ within the metastable regime. This means  that the variance of the activity, $\Delta^2 k(t)$, is inversely proportional to $t$.  If the metastable regime is long enough, the fluctuations of activity around its mean value, $N\kappa\left(m_{1,2}+\frac{1}{2}\right)$, are small and trajectories can be classified as ``active'' (large $k$) or ``inactive'' (small $k$). Note that when an initial state evolves at the beginning of the metastable regime into a mixture of two eMSs, $p_1 \tilde\rho_1+p_2 \tilde\rho_2$, the activity statistics is determined mostly by the contribution from the metastable regime, and thus emission records are a mixture of the active and inactive records.  

Now consider a \emph{coarse graining} in time of the photon emission records, where an emission record is divided into time bins of size ${t}_{\text{bin}}$ and the emission record is replaced by the activity of time bins. It is known that for intermittent dynamics the time-bin activity distribution can be bimodal~\cite{Ates2012Ising}. From the arguments above it follows that for a ${t}_{\text{bin}}$ well inside the metastable regime that is long enough, the probability distribution of activity is indeed bimodal with inactive dynamics corresponding to magnetisation of the ferromagnetic phase $m_1\approx-0.5$ and the active dynamics corresponding to the paramagnetic phase $m_2\approx 0$, which is visible in Fig.~\ref{IntTraj}\textcolor{blue}{(a)}.  This demonstrates that the periods of active and inactive dynamics (so called dynamical phases~\cite{Juan2010ThermoTraj}) correspond to directly to two metastable phases.     

We now discuss the relation between intermittency and the effective dynamics. For times after initial relaxation $t\gg\tau'$ the system dynamics $\rho(t)$ can be obtained as an average over classical trajectories generated by $\mathcal{L}_\text{eff}$, \eqref{EffectiveMasterOp}, between two metastable phases. Those classical trajectories are ergodic, i.e. time-averaged state in a trajectory approaches the steady state $\rho_{\rm ss}$ for $t\rightarrow\infty$, which requires the ratio of time spent in each of the metastable phases to be given by $p_{1}^{\rm ss}/p_{2}^{\rm ss}=-{c}_{2}^{\rm min}/{c}_{2}^{\rm max}$, cf.\ Eq.~\eqref{EffectiveMasterOp}. On the other hand, we know that the emission record coarse-grained in time consist of active and inactive periods with small fluctuations. As the activity of a long records approaches the emission rate of the steady state, $\left\langle{k}\right\rangle_{\rm ss}=\mu_{\rm ss}=N\kappa(m_{\rm ss}+\frac{1}{2})$, the probabilities of the inactive and active dynamics are again given (approximately) by $p_1^{\rm ss}$ and $p_2^{\rm ss}$ (see Fig.~\ref{IntTraj}\textcolor{blue}{(a)}). Finally, the average length of active/inactive dynamics in coarse-grained records cannot exceed the longest timescale $\tau$ of the model. This yields a direct relation between the classical trajectories and the coarse-grained intermittent emission records, see Fig.~\ref{IntTraj}\textcolor{blue}{(b-c)}.

The classical dynamics generated by $\mathcal{L}_\text{eff}$ can be related back to the double-well model discussed in Sec. \ref{MetaTheory}, with the sketches of the relevant double well potential shown in the insets of Fig.~\ref{IntTraj}\textcolor{blue}{(b-c)}. The smaller spectral gap $-{\lambda}_{2}$ for (c), leads to longer periods spent in two phases, analogous to a higher barrier between the wells, and results in a longer equilibration time. When the difference in probabilities between the metastable phases in $\rho_{\rm ss}$, $|p_{1}^{\rm ss}-p_{2}^{\rm ss}|=|{c}_{2}^{\rm max}+c_2^{\rm min}|/\Delta c_2$, is larger (b) the times spent in the two states differ more, analogously to how the bigger difference in free energies of the two wells makes time spent in the well of higher  free energy shorter and hence less of the equilibrium state is supported there. 
 
The average over the ensemble of quantum trajectories also gives the dynamics of the system state $\rho(t)$. It is not generally known, if those trajectories are ergodic, i.e. whether each time-averaged trajectory asymptotically gives the steady state. If that was the case, as the supports of metastable phases are approximately disjoint~\citep{Kasia2016Meta}, one could expect the quantum trajectories to be ergodic also inside those supports for $\tau'\ll t\ll \tau$, which would lead to the classical trajectories being coarse-graining of quantum ones.  By considering the magnetisation of a quantum trajectory, due to its relation to the photon emission activity, one indeed recovers the effective dynamics up to fluctuations within the $\tau'$ timescale, see Fig.~\ref{IntTraj}\textcolor{blue}{(b-c)}.

\section{Conclusions}

We have applied the general theory of metastability developed recently in Ref.~\citep{Kasia2016Meta} to the dynamics of an open many-body quantum system, namely the open quantum Ising model with transverse field in the presence of site decay, a model which could potentially be realised using Rydberg atoms.   For this system the origin and properties of the metastable dynamics is particularly transparent: in a region of parameter space where interactions and transverse field compete there is time separation in the dynamics, such that the two leading eigenmodes of the master operator split from the rest.  This means that the metastable manifold to which an initial state decays at short times is a one-dimensional simplex given by the convex combinations of two extreme metastable states.  These two states represent the two metastable phases of the problem, an almost pure state of negative magnetisation in the $z$-direction, and one mixed state of almost vanishing magnetisation.  We have also shown that the dynamics of this model at long times can be understood in terms of an effective classical dynamics between two phases inside the metastable manifold, which converges to the unique stationary state asymptotically.  

We have also shown that the short time behaviour and the existence of the long-lived metastable states are qualitatively captured by a straightforward mean-field approximation to the dynamics, very much in the spirit of what one does in slowly relaxing classical systems.  We have also discussed how metastability is directly related to fluctuating dynamics in individual realizations of the system evolution, specifically in the form of intermittency of quantum jump trajectories.   The physical ingredients that give rise to metastability in the model studied here are those that could also be responsible for the system having a true stationary state phase transition in the thermodynamic limit.  But just like in the case of analogous classical problems, the dynamics we uncovered (separation of timescales, transient relaxation to long-lived states, intermittency due to switching between metastable phases) is not dependent on that transition being present, as metastability would occur in systems where the transition is avoided either through finite size or due to other kind of fluctuations. 
The concrete application of the ideas of Ref.~\citep{Kasia2016Meta} to a many-body system should pave the way for further studies of metastability in other systems with even richer emergent collective dynamics.

\acknowledgments
The research leading to these results has received funding from the European Research Council under the European Union's Seventh Framework Programme (FP/2007-2013) / ERC Grant Agreement No. 335266 (ESCQUMA). We also acknowledge financial support from EPSRC Grant no.\ EP/M014266/1 and from the H2020-FETPROACT-2014 Grant No. 640378 (RYSQ).


\begin{thebibliography}{39}
\bibitem [{Jor()}]{JordanBlock}%
  \BibitemOpen
\bibfield  {journal} {  }\href@noop {} {}\bibinfo {note} {It is possible that
  the master operator may not be completely diagonalizable, in which this case
  its simplest form is the Jordan normal form. Under exponentiation this gives
  rise to a polynomial dependence on the time, however this is accompanied by
  the usual exponential evolution. The part of the spectrum we are concerned
  with is assumed to be diagonalizable (as is the case for this model) and any
  polynomial evolution is thus dominated by the decaying exponentials on the
  relevant time scales.}\BibitemShut {Stop}%
\end{thebibliography}
\end{document}